\documentclass[reprint,prb,aps,amsmath,amssymb,floatfix,superscriptaddress,longbibliography]{revtex4-2}

\usepackage{mathtools,amsmath}
\usepackage{graphicx} 
\usepackage[breaklinks=true,colorlinks,citecolor=teal,linkcolor=teal,urlcolor=teal]{hyperref}
\usepackage{physics}
\usepackage{siunitx}
\usepackage{bbold}
\usepackage{soul}
\usepackage{bm}
\usepackage{multirow}
\usepackage[normalem]{ulem}
\usepackage{times}
\usepackage{enumerate}  
\usepackage{float}
\usepackage{amssymb}
\usepackage{xcolor} 
\usepackage{booktabs}

\newcommand{\kBT}{k_{\rm B}T}

\begin{document}
\title{Making the Virtual Real:  Measurement-Powered Tunneling Engines}
\author{Rafael S\'anchez}
\affiliation{Departamento de F\'isica Teorica de la Materia Condensada, Universidad Aut\'onoma de Madrid, 28049 Madrid, Spain\looseness=-1}
\affiliation{Condensed Matter Physics Center (IFIMAC), Universidad Aut\'onoma de Madrid, 28049 Madrid, Spain\looseness=-1}
\affiliation{Instituto Nicolas Cabrera, Universidad Aut\'onoma de Madrid, 28049 Madrid, Spain\looseness=-1}
\author{Alok Nath Singh}
\affiliation{Department of Physics and Astronomy, University of Rochester, Rochester, NY 14627, USA}
\affiliation{Institute for Quantum Studies, Chapman University, Orange, CA 92866, USA}
\author{\\Andrew N. Jordan}
\affiliation{Institute for Quantum Studies, Chapman University, Orange, CA 92866, USA}
\affiliation{Schmid College of Science and Technology, Chapman University, Orange, CA, 92866, USA}
\affiliation{Department of Physics and Astronomy, University of Rochester, Rochester, NY 14627, USA}
\affiliation{The Kennedy Chair in Physics, Chapman University, Orange, CA 92866, USA}
\author{Bibek Bhandari}
\affiliation{Institute for Quantum Studies, Chapman University, Orange, CA 92866, USA}

\begin{abstract}
Quantum tunneling allows electrons to be transferred between two regions separated by an energetically forbidden barrier. Performing a position measurement that finds a particle in the barrier forces the tunneling electrons to transition from having a classically forbidden energy to an energy above the barrier height. We exploit this effect to define quantum tunneling engines that can use the unconditioned detection of virtually occupied states as a resource for power generation and cooling. Leveraging energy exchange with the detector, we show that the device can operate in a hybrid regime, enabling simultaneous cooling and power generation. Furthermore, we demonstrate measurement-assisted autonomous refrigeration and {\em checkpoint} cooling driven purely by a thermal bias, without the need for an applied potential. We also find a {\em purification by noise} effect when the measurement drives the system into a stationary dark state. These results underscore the intriguing dual role of measurement as a thermodynamic resource and a dark state generator. 
\end{abstract}

\maketitle

\section{Introduction}
\label{sec:intro}

The act of measuring has dramatic consequences in quantum systems~\cite{andrew_book,wiseman_book}. After decades of thinking about the unavoidable backaction due to measurement as a problem~\cite{hussein_monitoring_2014,sankar_backaction_2025}, the scope has changed in recent years to try to use it to our advantage. 
Particular attention has been put to connections with quantum thermodynamics~\cite{binder:2018,alexia_initiative}, where the detection apparatus can serve as a resource to perform useful operations on the system--such as work production~\cite{elouard_efficient_2018,elouard_interaction_2020,bresque_twoqubit_2021}, refrigeration~\cite{ferreira_transport_2024,bhandari2020continuous,bhandari2023measurement}, quantum elevator \cite{elouard_efficient_2018,elouard_interaction_2020,jordanquantum2020}, quantum battery \cite{gherardini2020stabilizing,mitchison2021charging,zhang2024local}, and information-driven devices such as the Szilard engine~\cite{mohammady2017quantum,koski:2014} and quantum Maxwell demons~\cite{yanik2022thermodynamics,erdman2024artificially}. Experiments have been performed demonstrating the use of feedback-based detection as a quantum Maxwell demon~\cite{cottet:2017,masuyama:2018}, showcasing how measurement and feedback can serve to control entropy dynamics in quantum systems.

\begin{figure}[b]
\includegraphics[width=\linewidth]{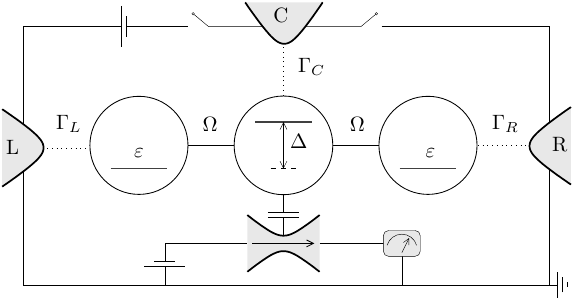}
\caption{\label{fig:scheme_3t}\small Triple quantum dot engine fueled by a quantum point contact detector. Each dot is coupled to a different reservoir $l$=L,C,R via tunneling rates $\Gamma_l$. Electrons tunneling between the left and right quantum dots with energies $\varepsilon$ are detected when virtually occupying the central one detuned by an energy $\Delta$ and absorbed by reservoir C. Interdot tunneling is given by $\Omega$. The electrochemical potential of C can be either tuned to have a power-generating engine or a quantum state purifier, or grounded to have a refrigerator.}
\end{figure}

Position monitoring localizes the wavefunction within the measurement-probed region, thereby altering particle dynamics~\cite{elouard_efficient_2018,mackrory_reflection_2010}. This effect persists even when the detection zone is only virtually occupied, for instance, during quantum tunneling through a potential barrier, where the detector supplies the energy required to access otherwise inaccessible higher-energy states~\cite{singh2025capturing,romito_weak_2014,zilberberg2014,zilberberg2019}. In this work, we propose a triple quantum dot (TQD) array-based setup~\cite{gaudreau_stability_2006,schroer_electrostatically_2007,hsieh_physics_2012} (see Fig.~\ref{fig:scheme_3t}) to use this mechanism for practical thermodynamic operations, including electric power generation and cooling. A potential barrier is engineered in the system by electrostatically gating the central quantum dot off-resonance, creating a tunable platform to explore measurement-driven energy conversion.

When the central dot is strongly detuned from the two external ones, hopping between left and right dots (fed by two electronic reservoirs) occurs via virtual tunneling transitions~\cite{ratner_bridge_1990,amaha_resonance_2012,busl_bipolar_2013,braakman_long_2013,sanchez_longrange_2014,superexchange,contrerasPulido_dephasing_2014,tormo-Queralt_novel_2022,aizawa_dynamics_2024}. However if the central dot is coupled to a charge detector, modeled here as a quantum point contact~\cite{field_measurements_1993,gurvitz_measurements_1997,buks_dephasing_1998,gustavsson:2006,fujisawa:2006,ubbelohde:2012,kung_irreversibility_2012,hofmann:2016}, these transitions can be detected in the act~\cite{singh2025capturing}, resulting in the {\it actual} occupation of the central dot. A collector reservoir C absorbs these excitations, realizing the measurement-induced local charge accumulation into a finite current.

The TQD device operates autonomously, leveraging measurement backaction while discarding detection outcomes (i.e., no feedback is applied). In doing so, it extends the class of autonomous quantum dot engines proposed in recent works~\cite{sothmann:2015,benenti:2017,cangemi_quantum_2024,guzman_key_2024,balduque_quantum_2025}, 
which utilize thermodynamic resources to achieve functionality. Notably, parallels can be drawn to proposals that re-purpose heat for quantum information tasks~\cite{brask:2015njp,tavakoli:2018}. This analogy underscores a critical insight: the measurement process facilitates heat exchange between the detector and the device, enabling performance characterization via the ratio of useful output (e.g., power generation or cooling) to heat transfer.
This mechanism is distinct from transport generation driven by direct energy exchange with the current in a nonequilibrium conductor~\cite{khrapai_doubledot_2006,harbusch_phonon_2010,bischoff:2015,keller:2016}.

The triple quantum dot’s coherent properties, specifically, the hybridization of its three atomic-like states into extended molecular-like superpositions, enable additional functionality. In left-right (L-R) symmetric configurations, one such superposition excludes the central dot entirely due to destructive interference, decoupling it from transport into reservoir C. These ``dark states", analogous to those proposed for coherent population trapping under nonequilibrium conditions~\cite{brandes_current_2000,brandes_coherent_2005,michaelis_allelectronic_2006} or entanglement generation~\cite{sanchez_dark_2013,zhou_quantum_2024}, leave distinct transport signatures~\cite{emary_dark_2007,niklas_fano_2017,kostyrko_symmetry_2009,dominguez_electron_2010,dominguez_phonon_2011,superexchange,donarini_coherent_2019}. Unlike all other states in the system, the dark state remains unaffected by the detector (which precisely only measures the occupation of the center dot). Under specific parameter regimes, its occupation becomes dynamically stable, even with all reservoirs in equilibrium, allowing us to steer the system into a steady dark state, a phenomenon we term ``purification by noise".

The remainder of the paper is organized as follows: in Sec.~\ref{sec:model}, we present the theoretical model, the dynamic equations, and the thermodynamic quantities. The performance of the system as a quantum thermodynamic engine or as a steady state purifier is discussed in Secs.~\ref{sec:thermo_res} and \ref{sec:purif}, respectively, with conclusions presented in Sec.~\ref{sec:conc}.

\section{Triple quantum dot system}
\label{sec:model}

We consider a linearly aligned triple quantum dot (TQD) system, where each dot is weakly coupled to a distinct electronic reservoir, $l = \{{\mathrm{L}, \mathrm{C}, \mathrm{R}}\}$, with an electrochemical potential $\mu_l$ and a temperature $T_l$, as illustrated in Fig.~\ref{fig:scheme_3t}. Unless stated otherwise, all reservoirs are maintained at the same temperature $T$ and $\mu_L=\mu_R=\mu$. We label the dots the same way as the reservoirs to which they are coupled. The central dot, C, is tunnel coupled to the other two dots by a hopping strength $\Omega$, with no direct coupling between the outer dots. The central dot is additionally coupled to a quantum point contact (QPC) charge detector. 

We assume that the system is in the strong Coulomb blockade regime~\cite{vanderWiel_electron_2002}, where Coulomb interactions prevent the TQD from hosting more than one electron at a time. In this case, the spin degree of freedom is irrelevant, so we will omit it for simplicity. The Hamiltonian of the system is
\begin{align}
\label{ham}
\hat{H}_{\text{TQD}} = \sum_{l}\varepsilon_{l}\hat{d}^{\dagger}_{l}
\hat{d}_{l}^{}
-\sum_{l\neq \rm C}\left(\Omega\hat{d}^{\dagger}_{l}\hat{d}_{\rm C}^{}+\text{H.c.}\right),
\end{align}
where the operator $\hat{d}_{l}^{}$ creates an electron in quantum dot $l$, and $\varepsilon_l$ is the energy of the occupied quantum dot. The configuration space is restricted to four states: the empty $|0\rangle$, and the singly occupied ones $|l\rangle=\hat{d}_l^{\dagger}|0\rangle$.

The reservoirs and their coupling to the TQD are described by the Hamiltonian terms $\hat{H}_{\text{res}} = \sum_{l,k}\varepsilon_{lk}\hat{c}^{\dagger}_{lk}\hat{c}_{lk}$ and $\hat{H}_{\text{tun}} = \sum_{l,k}\gamma_{l}\hat{d}^{\dagger}_{l}
\hat{c}_{lk}+\text{h.c.}$, where $\hat{c}^{\dagger}_{lk}$ creates a free electron of momentum $\hbar k$ in reservoir $l$. The coupling constant will define the tunneling rate via Fermi's golden rule: $\Gamma_l=2\pi\hbar^{-1}|\gamma_l|^2\nu_l$, with $\nu_l$ being the density of states in reservoir $l$. We will monitor the charge of the central dot by coupling it to a QPC with a detection rate $\gamma$ to distinguish the state $\ket{0}$ from $\ket{C}$ in the central dot.

We are interested in a configuration where the outer dots have the same energy, $\varepsilon_{\rm L}=\varepsilon_{\rm R}=\varepsilon$, so the electron transfer along the chain conserves energy. They are, however, separated by a detuned central dot with $\varepsilon_{\rm C}=\varepsilon+\Delta$, forming a bridge~\cite{ratner_bridge_1990}. As long as $\Delta\gg\Omega$, the central dot does not hybridize with the other two, and hence its occupation via interdot hopping is avoided. Nevertheless, direct tunneling between the outer dots is possible via virtual transitions involving C. This process has been detected in the form of narrow resonances in the current~\cite{busl_bipolar_2013,sanchez_longrange_2014} or via charge monitoring~\cite{braakman_long_2013}, see also Ref.~\onlinecite{amaha_resonance_2012}. This way, the TQD is a discrete version of a tunnel barrier, with the detuning of the central dot determining the height of the barrier. A perturbative expansion gives an effective coupling, $\Omega_{\rm eff}=\Omega^2/\Delta$, for the virtual tunneling~\cite{ratner_bridge_1990,braakman_long_2013}, see also Ref.~\onlinecite{girvinbook}.

\subsection{Master Equation}
\label{sec:mastereq}
To have a thermodynamically consistent description, we use the system global basis obtained by diagonalizing $\hat{H}_{\rm TQD}$ through the change of basis given by
\begin{gather}
\begin{aligned}
\label{eq:1e_eigst}
|D\rangle&=(|L\rangle-|R\rangle)/\sqrt{2},\\
|\pm\rangle&=\theta_{\Omega\pm}(|L\rangle+|R\rangle)-\theta_{\alpha\pm}|C\rangle,
\end{aligned}
\end{gather}
with $\theta_{\Omega\pm}=\Omega/{\cal N}_\pm$, $\theta_{\alpha\pm}=\alpha_\pm/{\cal N}_\pm$, $\alpha_\pm\equiv(\Delta\pm\chi)/2$, $\chi=\sqrt{\Delta^2+8\Omega^2}$, and ${\cal N}_\pm=\sqrt{2\Omega^2+\alpha_\pm^2}$. 
Here, the central dot does not contribute to the superposition $|D\rangle$, whose labeling stands for a ``dark'' state~\cite{michaelis_allelectronic_2006}.
The eigenenergies are $E_{\rm D}=\varepsilon$ and $E_\pm=\varepsilon+\alpha_\pm$. Note the splitting, $E_+-E_-=\chi$.
With this notation, the tunneling rates $W_{ji}^l$ for transitions $|i\rangle\to|j\rangle$ involving transitions associated with reservoir $l$ can be calculated using Fermi's golden rule from the matrix elements ${|}\bra{j}\hat{H}_{\rm tun}\ket{i}{|}^2$ as 
\begin{gather}
\begin{aligned}
W_{\pm0}^{\rm L/R}&=\Gamma_{\rm L/R}\theta_{\Omega_\pm}^2f(E_\pm-\mu_{\rm L/R}),\\
W_{\rm D0}^{\rm L/R}&=\frac{1}{2}\Gamma_{\rm L/R}f(E_{\rm D}-\mu_{\rm L/R}),\\
W_{\pm0}^{\rm C}&=\Gamma_{\rm C}\theta_{\alpha\pm}^2f(E_\pm-\mu_{\rm C}),\\
W_{\rm D0}^{\rm C}&=0,
\end{aligned}
\end{gather}
with the reversed transitions obtained by replacing the Fermi function $f(E)=1/[1+\exp(E/k_{\rm B}T)]$ by $1-f(E)$.

To write the Lindbladian associated with the measurement $(\hat L_{\rm M})$, we write $|C\rangle=\beta_-|-\rangle-\beta_+|+\rangle$, with $\beta_\pm={\cal N}_\pm/\chi$,~\footnote{Note that $\beta_+=\theta_{\alpha+}$ and $\beta_-=-\theta_{\alpha-}$, so we fulfill $\langle+|C\rangle=-\beta_+=-\theta_{\alpha+}$ and $\langle-|C\rangle=\beta_-=-\theta_{\alpha-}$, as we expect from Eq.~\eqref{eq:1e_eigst}.}
such that
\begin{align}
\hat{L}_{\rm M}&=\sqrt{\gamma}|C\rangle\langle C|=\sqrt{\gamma}\sum_{i,j=\pm}ij\beta_i\beta_j|i\rangle\langle j|.
\end{align}

In the weak system-reservoir coupling regime, $\Gamma_l\lesssim \Omega\ll \Delta$, where we can neglect hopping-induced contributions of the off-diagonal elements~\cite{potts:2021,correa24testing,bibekgreen}, the following master equations give the evolution of the system occupations:
\begin{align}
\label{eq:p00}
\dot\rho_{00}&=\sum_{l,\lambda}\left(W_{0\lambda}^l\rho_{\lambda\lambda}-W_{\lambda0}^l\rho_{00}\right),\\
\label{eq:p++}
\dot\rho_{++}&=\sum_l\left(W_{+0}^l\rho_{00}-W_{0+}^l\rho_{++}\right)+\gamma\beta_+^2\beta_-^2(\rho_{--}-\rho_{++})\nonumber\\
&-\gamma\beta_+\beta_-\Lambda X_{},\\
\label{eq:p--}
\dot\rho_{--}&=\sum_l\left(W_{-0}^l\rho_{00}-W_{0-}^l\rho_{--}\right)+\gamma\beta_+^2\beta_-^2(\rho_{++}-\rho_{--})\nonumber\\
&+\gamma\beta_+\beta_-\Lambda X_{},\\
\label{eq:pDD}
\dot\rho_{\rm DD}&=\sum_l\left(W_{\rm D0}^l\rho_{00}-W_{\rm 0D}^l\rho_{\rm DD}\right),
\end{align}
with the index $i=\pm$ and $\lambda\in\{\rm D,\pm\}$ accounting for the single-particle states~\eqref{eq:1e_eigst}, and where we have defined 
$\Lambda\equiv\beta_+^2-\beta_-^2$.
We split the coherences as $\rho_{+-}=X_{}+iY_{}$ for convenience, which evolve as
\begin{align}
\label{eq:X}
\dot{X}_{}&=\frac{\chi}{\hbar}Y_{}-\frac{1}{2}\left[\sum_{l,i}W_{0i}^l+\gamma\left(\beta_+^2{-}\beta_-^2\right)^2\right]X_{}\nonumber\\
&-\frac{\gamma}{2}\beta_+\beta_-\Lambda(\rho_{++}-\rho_{--}),
\\
\label{eq:Y}
\dot{Y}_{}&=-\frac{\chi}{\hbar}X_{}-\frac{1}{2}\left[\sum_{l,i}W_{0i}^l+\gamma\left(\beta_+^2{+}\beta_-^2\right)^2\right]Y_{}.
\end{align} 
Note that the effect of the detector enters the dynamics of the populations and their coupling to the coherences via the imbalance $\rho_{++}-\rho_{--}$ in Eqs.~\eqref{eq:p++}, \eqref{eq:p--} and \eqref{eq:X}. In particular, in configurations with stationary states satisfying $\rho_{++}=\rho_{--}$, the system is dynamically decoupled from the detector.

\subsection{Highly Detuned central dot}
\label{sec:limits}
The occupation of the central dot is negligible in the absence of the detector when the central dot is highly detuned compared to the outer dots, i.e, $\Delta\gg\Omega,k_{\rm B}T,|\varepsilon-\mu|$. In this case, we can approximate $\alpha_+\approx\Delta$, $\alpha_-\approx-2\Omega^2/\Delta$, $\theta_{\Omega+}\approx\Omega/\Delta$, $\theta_{\Omega-}\approx1/\sqrt{2}$, $\theta_{\alpha+}\approx1$ and $\theta_{\alpha-}\approx-\sqrt{2}\Omega/\Delta$, such that $E_+\approx\varepsilon+\Delta$ and $E_-\approx\varepsilon$. In the absence of potential bias, we define $f(E_i-\mu_l)=f(E_i) \equiv f_i$, such that $f_+\to0$. Further, since $\Omega\ll \Delta$, the charge current into reservoir C mostly depends on the occupation of the central dot and is given by $I_{\rm C}\approx\Gamma_{\rm C}\rho_{\rm CC}$, where the occupation probability of the central dot can be obtained from Eq.~(\ref{eq:1e_eigst}) as
\begin{equation}
\rho_{\rm CC}=\theta_{\alpha+}^2\rho_{++}+\theta_{\alpha-}^2\rho_{--}.
\end{equation}
Further considering $\Gamma_{ l}=\Gamma$ and $\gamma \to 0$ corresponding to a very weak measurement limit, we can expand the occupation probability of the central dot to the leading order in $\Omega/\Delta$. We get
\begin{equation}
\label{eq:rcc}
\rho_{\rm CC}\approx\rho_{\rm CC}^{(0)}\left(1+\frac{\gamma}{\Gamma}\right)+{\cal O}(\gamma^2),
\end{equation}
with the residual occupation of the central dot in the absence of the detector
\begin{equation}
\rho_{\rm CC}^{(0)}\equiv\frac{2f_-(1-f_{\rm D})}{1-f_-f_{\rm D}}\frac{\Omega^2}{\Delta^2}.
\label{eq:rcc0}
\end{equation}
In the low-temperature limit \footnote{Low temperature here means that $\varepsilon-\mu\gg k_{\rm B}T$, i.e., the same limit corresponds to a finite temperature, but energy states well below the chemical potential.}, when $\varepsilon<\mu$, we can approximate $f_-\to1$ and $f_{\rm D}\to1$. As $E_-<E_{\rm D}$, we will consider $f_-\to1$ first, getting $\rho_{\rm CC}^{(0)}\to2\Omega^2/\Delta^2 = \langle C|-\rangle^2 $. Hence,  this contribution does not lead to any transport to reservoir C since $E_-$ is well below the Fermi level of reservoir C for $\epsilon<\mu$.
Further, the master equations would result in a bistable stationary state, as both $|-\rangle$ and $|D\rangle$ have almost identical eigenenergies, $E_-\approx E_{\rm D} = \epsilon$.

If we consider, $f_-\to1$ and $f_{\rm D}\to1$, directly on Eq.~\eqref{eq:rcc0} without further expansion on $\Omega/\Delta$, we get $\rho_{\rm CC}\to0$, independently of $\gamma$, i.e., although the detector tries to populate the central dot, its population is quickly transferred to the outer dots. In other words, even a very small coupling to the detector will wash out the contribution of $|C\rangle$ to the state $|-\rangle$, leading to a stationary occupation of the dark state. In this case, the density matrix reduces to a pure state density matrix ($\rho=|D\rangle\langle D|$)--the detector purifies the system. We will postpone the discussion on purification by detection to Sec.~\ref{sec:purif} and investigate the finite temperature applications of the device in the next sections. 

\subsection{Detection-Induced Transport}
\label{sec:transport}

\begin{figure}[t]
\includegraphics[width=\linewidth]{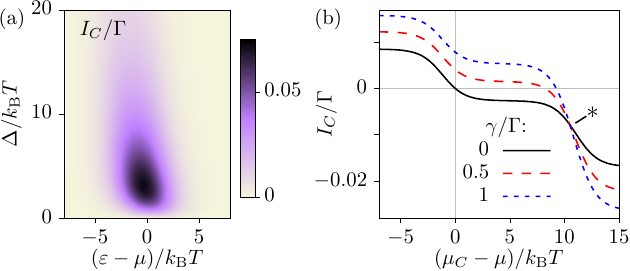}
\caption{\label{fig:current_3t_gating}\small (a) Zero bias generated current $I_{\rm C}$ in units of $e\Gamma$ as a function of the position of the outer quantum dot levels, $\varepsilon$ and the splitting of the central dot with respect to them, $\Delta$, with $\gamma=5\Gamma$, and (b) as a function of the bias $\mu_{\rm C}-\mu$ for different detection strengths, with $\varepsilon-\mu=-k_{\rm B}T$ and  $\Delta=10k_{\rm B}T$. Other parameters: $\Omega=k_{\rm B}T$, $\Gamma=0.1k_{\rm B}T/\hbar$, and $\mu_L=\mu_R=\mu=0$. The black star in panel (b) denotes the potential bias where the current becomes independent of the detector strength. A finite measurement-induced current is observed even in the absence of potential bias ($\mu_{\rm C} = \mu$). 
}
\end{figure}
Particle currents can be calculated using the occupation probabilities obtained from the stationary solution, $\dot\rho=0$, as
\begin{equation}
I_l=\sum_{\lambda=\{\rm D,\pm\}}\left(W_{0\lambda}^l\rho_{\lambda\lambda}-W_{\lambda0}^l\rho_{00}\right),
\end{equation}
defined as positive when they flow into a reservoir. The first term on the right-hand side gives the incoming processes 
(from the TQD to the reservoir $l$) whereas the second term gives the outgoing processes (from the reservoir $l$ to the TQD). In Fig.~\ref{fig:current_3t_gating} (a), we show the current generated by the measurement process (with all three reservoirs being in equilibrium) as a function of the position of the quantum dot levels. The current takes its maximum when the detuning is of the order of the inter-dot hopping, $\Omega$, where the occupation of the central dot is not forbidden. A long tail toward the regime $\Delta\gg\Omega$ exhibits the current induced by the detection of electrons in the act of tunneling~\cite{singh2025capturing}. We note that the current vanishes when both the measurement and potential bias are absent, as expected for an isolated system in equilibrium, see the solid black curve at $\mu_{\rm C} = \mu$ in Fig.~\ref{fig:current_3t_gating}(b). However, a finite current emerges at zero potential bias when a measurement is applied (red dashed and blue dotted curves at $\mu_{\rm C} = \mu$).

In the following, we will focus on the regime where the central dot is highly detuned, i.e., $\Delta\gg k_{\rm B }T$. In this case, the current in reservoir C in response to an electrochemical potential bias, $\mu_{\rm C}-\mu$, shows a double-step behavior, see Fig.~\ref{fig:current_3t_gating}(b). The double step for $\gamma=0$ is consistent with having a triple dot: steps appear around the crossings of the quantum dot levels with the chemical potential of the reservoir. Further, the small thermal broadening enforced by the condition $\Delta\gg k_{\rm B }T$ ensures the appearance of the steps. Interestingly, at a particular potential bias ($\mu^*$, pointed by an asterisk in Fig.~\ref{fig:current_3t_gating}(b)), the current becomes independent of the measurement strength. This happens close to the condition where $E_+-\varepsilon=\mu_{\rm C}$, i.e., when the large detuning of the central dot is compensated by the chemical potential of the central reservoir. 
At this condition, the system also fulfills the condition $\rho_{++}=\rho_{--}$, effectively decoupling the detector from the quantum dot system (recall the dependence of master equations \eqref{eq:p00}-\eqref{eq:Y} on the measurement strength $\gamma$ discussed at the end of Sec.~\ref{sec:mastereq}).

\begin{figure}[t]
\includegraphics[width=\linewidth]{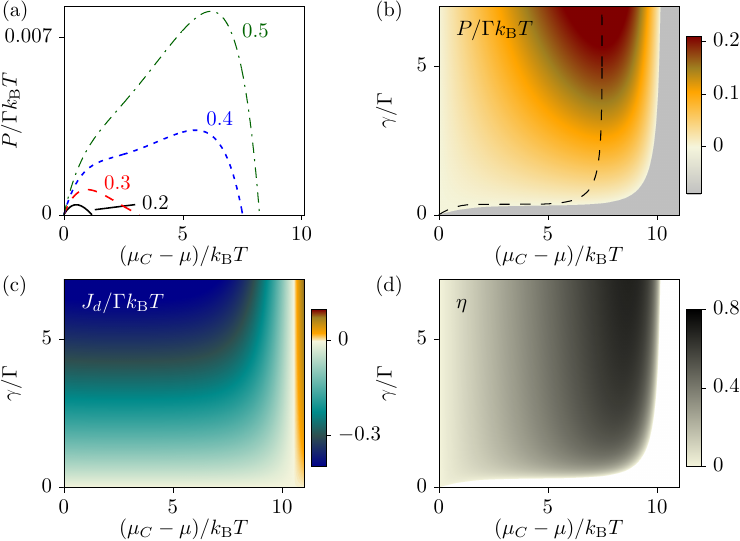}
\caption{\label{fig:current_3t}\small  (a) Generated power, $P$, as a function of $\mu_{\rm C}$ for different detection strengths $\gamma/\Gamma$. Contour plots of (b) $(P)$, (c) the heat current exchanged with the detector, $J_{\rm d}$, and (d) the efficiency, $\eta$, as functions of the measurement strength $\gamma$ and the potential bias $\mu_{\rm C} - \mu$. Negative values in (b), indicating dissipated power, are shown in gray for clarity. The same parameters as in Fig.~\ref{fig:current_3t_gating} are used.}
\end{figure}

The heat current flowing into the reservoir $l$ is given by
\begin{equation}
J_l=\sum_\lambda(E_\lambda-\mu_l)\left(W_{0\lambda}^l\rho_{\lambda\lambda}-W_{\lambda0}^l\rho_{00}\right),
\end{equation}
with the generated electrical power being
\begin{equation}
P=I_{\rm C}(\mu_{\rm C}-\mu).
\end{equation}
The system works as an engine when $P>0$. In contrast to Ref.~\cite{singh2025capturing}, we consider a strongly hybridized triple quantum dot system described in the global eigenbasis, where a local measurement induces nontrivial energy exchange, manifested by a finite heat exchange with the measurement probe, $J_{\rm d} = \Tr{\hat H_{\rm TQD}\hat{L}_{\rm M}\hat \rho_{\rm TQD}} \neq 0$, where $\hat \rho_{\rm TQD}$ is the TQD reduced density matrix. Using the conservation of energy, we get the heat current exchanged with the detector~\cite{bhandari2020continuous}:
\begin{equation}
J_{\rm d}=-(P+J_{\rm L}+J_{\rm R}+J_{\rm C}).
\end{equation}
We use the same sign convention for all heat currents: $J_l$ is positive when heat current flows into the reservoir, $J_{\rm d}$ is negative when flowing out of the detector. 
With the above definition for heat current and power generation, and neglecting dissipation in the detector, we can define the engine efficiency as
\begin{equation}
\eta=\frac{P}{-J_{\rm d}}. 
\end{equation}
We ignore the heat dissipated in the detector due to its own operation, as it has no impact on the TQD.

\section{Measurement as a thermodynamic resource}
\label{sec:thermo_res}
Let us explore how the measurement backaction, rather than traditional thermodynamic biases such as voltage or thermal differences, can be exploited as a resource for quantum thermodynamic operations: A) Heat engines and refrigerators, B) hybrid operations with simultaneous power generation and refrigeration, and C) autonomous refrigeration.

\subsection{Heat Engines and Refrigerators}
\label{sec:engines}
We first analyze the operation of the measurement-powered TQD system as a heat engine or a refrigerator. Figure \ref{fig:current_3t} shows the engine performance--generated power, heat current into the detector, and efficiency--as functions of the measurement strength ($\gamma$) and the applied potential bias. We will define the potential bias after which the system cannot be operated as a heat engine as ``stall voltage" (see the edge of the yellow contours followed by a gray region in Fig.~\ref{fig:current_3t}(b)).

In Fig.~\ref{fig:current_3t}(a), we plot the generated power of the system as a function of the applied potential bias for varying detector strengths. For all voltages, the generated power increases monotonically with $\gamma$, with the opposite being true for the absorbed heat $J_d$, see Figs.~\ref{fig:current_3t}(b) and \ref{fig:current_3t}(c).
We observe two well-defined regimes with different features: 
(i) For very weakly coupled detectors ($\gamma\ll\Gamma$), a peak at low voltages appears, whose maximum power and stall potentials increase with $\gamma$, mimicking the behaviour of quantum heat engines coupled to a thermal source~\cite{hotspots}. 
(ii) For larger couplings ($\gamma\gtrsim\Gamma$), a large response dominates whose maximum power and stall potentials are independent of $\gamma$, as shown in Fig.~\ref{fig:current_3t}(b). In this case, the voltage at maximum power is close to the corresponding stall voltage, a property which is beneficial for increasing the efficiency. In fact, $\eta$ saturates close to $0.8$ for the chosen parameters (see Fig.~\ref{fig:current_3t}(d)). For the same reason, the regions for maximal power and maximal efficiency coincide.

Notably, the heat current injected from the detector changes sign at a critical chemical potential $\mu^*$, where the charge current also becomes independent of $\gamma$, see Figs.~\ref{fig:current_3t_gating}(b) and \ref{fig:current_3t}(c). This marks a clear and effective decoupling of the detector from the system, which coincides with a transition in the occupation of the central dot, from virtual to real states, driven by direct injection from reservoir C at sufficiently large $\mu_{\rm C}-\mu$. This transition fundamentally alters the system’s charge and heat transport characteristics (see comparative analysis in the App.~\ref{app:vanishing}). The vanishing of $J_{\rm d}$ occurs when the system parameters satisfy $\rho_{++}=\rho_{--}$. When $\varepsilon\lesssim\mu$ [the region of interest, where power is generated, see Fig.~\ref{fig:current_3t}(c)], this roughly coincides with the chemical potential $\mu_{\rm C}$ crossing the energy $E_+$ (see Fig.~\ref{fig:3t_vanishinJd} in the App.~\ref{app:vanishing}). Here, the central dot transitions from virtual to real occupation due to interactions with reservoir C.

\subsection{Hybrid Operations}
\label{sec:hybrid}
\begin{figure}[t]
\includegraphics[width=.7\linewidth]{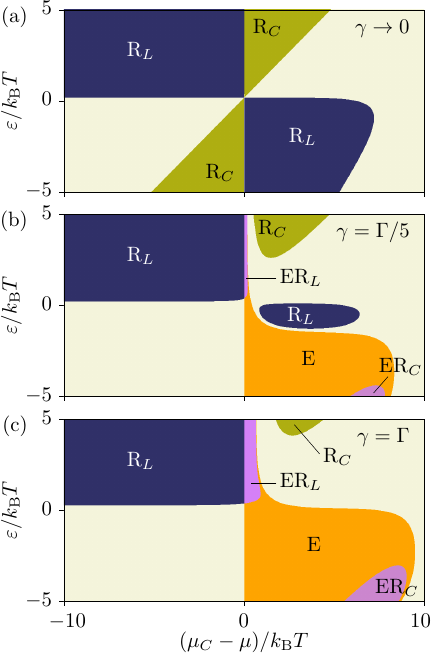}
\caption{\label{fig:hybridmap}\small Operations of the device for different couplings to the detector: (a) $\gamma\to0$, (b) $\gamma=\Gamma/5$ and (c) $\gamma=\Gamma$, with parameters as given in Fig.~\ref{fig:current_3t_gating}. The operational regimes are classified as follows: ${\rm R}_l$ denotes the region where the system operates as a refrigerator for reservoir $l$ $(J_l<0)$; E corresponds to power generation ($P > 0$); and ${\rm ER}_l$ identifies hybrid regions where refrigeration and power generation coexist.}
\end{figure}

The detector enables the system to operate as a hybrid engine, capable of simultaneously generating power and refrigerating either the outer reservoirs (L and R) or the central reservoir (C). This is a known feature of engines coupled to multiple reservoirs~\cite{entin:2015,manzano_hybrid_2020,tabatabaei_nonlocal_2022,lopez_optimal_2023,lu_multitask_2023} or work sources~\cite{hammam_exploting_2022,cavaliere_hybrid_2023,hammam_quantum_2024}. As illustrated in Fig.~\ref{fig:hybridmap}, this hybrid functionality is governed by tuning the chemical potential $\mu_{\rm C}$ and the quantum dot energy level $\varepsilon$, under the condition that the outer reservoirs share identical chemical potentials and temperatures. The operational regimes are classified as follows: ${\rm R}_l$ denotes the region where the system works as a refrigerator for reservoir $l$ (i.e., when the heat current $J_l<0$ being $T_l\leq T_{l'}$, for any $l'\neq l$), E corresponds to power generation ($P>0$), and ${\rm ER}_l$ denotes the hybrid regions where refrigeration and power generation coexist. Notably, the particle and heat currents in reservoirs L and R remain equal in the symmetric configuration we are considering ($\mu_L=\mu_R$, $T_L=T_R$).

For $\gamma = 0$, refrigeration in the ${\rm R}_{\rm L}$ and ${\rm R}_{\rm C}$ regions originates from Peltier cooling, see Fig.~\ref{fig:hybridmap}(a). Here, the voltage-driven particle current extracts heat from a reservoir whenever electrons enter/leave the reservoir while carrying energy above/below the reservoir chemical potential~\cite{benenti:2017}. When $\gamma\neq 0$, two key mechanisms alter the above behavior. First, the particle flow is dictated by the interplay between the bias $(\mu_{\rm C}-\mu)$ and the action of the detector. Second, electrons with previously forbidden energies ($\varepsilon + \Delta$) are pumped into reservoir C due to the detected virtual transitions resulting in the real occupation of the central quantum dot. The first mechanism modifies the ${\rm R}_{\rm L}$ and ${\rm R}_{\rm C}$ regions for $\varepsilon>0$, hybridizing the ${\rm R}_{\rm L}$ regime while shrinking ${\rm R}_{\rm C}$ (see Figs.~\ref{fig:hybridmap}(b) and \ref{fig:hybridmap}(c)). The second mechanism shifts the ${\rm R}_{\rm C}$ region for $\varepsilon<0$ to higher bias values, hybridizing it as well, and suppresses the ${\rm R}_{\rm L}$ region. This suppression occurs because detected electrons tunnel back into the system at elevated energies, reversing the heat extraction mechanism.

Furthermore, in the hybrid ${\rm ER_C}$ region, electrons are pumped into reservoir C below its chemical potential, an unconventional transport regime where both particle and heat currents are inverted. The power generation comes at the cost of decreased cooling of the reservoir C with increasing measurement strength. The hybrid operation regions (${\rm ER}_{\rm L}$ and ${\rm ER}_{\rm C}$) are broadened for larger detection strengths (see Fig.~\ref{fig:hybridmap}(c) for the case $\gamma=\Gamma$), however the region ${\rm R}_{\rm C}$ shrinks further.

\subsection{Autonomous and Checkpoint Refrigerators}
\label{sec:autonomous}

\begin{figure}[t]
\includegraphics[width=\linewidth]{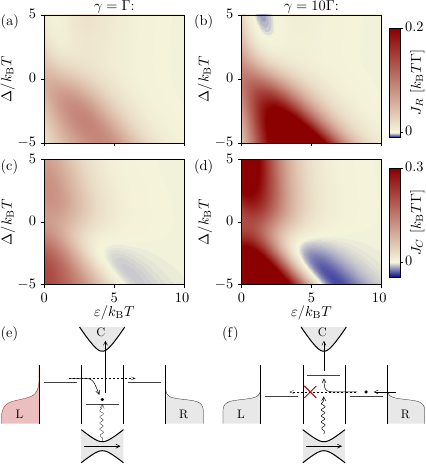}
\caption{Autonomous refrigerator: Heat currents in reservoirs (a), (b) $R$ and (c), (d) $C$ for different couplings to the detector: (a), (c) $\gamma=\Gamma$ and (b), (d) $\gamma=10\Gamma$, as functions of the position of the quantum dot levels. Parameters: $\Omega/\kBT=1$, $T_R=T_C=T$, $T_L=1.1T$, $\Gamma/\kBT=0.1$ and $\mu_L=\mu_R=\mu=0$. Schematic descriptions of the involved processes are shown in (e) for the absorption refrigeration and in (f) for checkpoint cooling.} 
\label{fig:autonomous}
\end{figure}
The detector is also able to induce cooling of a cold reservoir by employing a thermal bias, rather than a potential bias. In our setup, we investigate this process by cooling either the right or central reservoir, driven by a hot reservoir L and sustained by continuous measurement of the central quantum dot. This is done via two different mechanisms that distinguish the two main back-action effects: the energy exchange with the detector (for cooling C) and the selection of dynamical paths (for cooling R). 

The configuration follows Fig.~\ref{fig:scheme_3t}, with chemical potentials fixed at $\mu_{\rm L} = \mu_{\rm R} = \mu_{\rm C} = 0$, temperatures $T_{\rm C} = T_{\rm R}$, and $T_{\rm L}/T_{\rm C} = 1.1$, see Fig.~\ref{fig:autonomous}. We focus in configurations with $\varepsilon>\mu$, such that particles are preferably injected from reservoir L than from R. While the central and right reservoirs, being colder, typically exhibit positive heat currents for $\gamma=0$, continuous measurement enables parameter regimes where either reservoir is cooled. Note however that that these mechanisms do not require a temperature difference to be activated.

For $\gamma=\Gamma$, cooling is restricted to the central reservoir (no dark blue region in Fig.~\ref{fig:autonomous}(a)): for negative $\Delta$, out of all electrons injected from L, those that are detected are subsequently absorbed by reservoir C below $\mu_C$, as depicted in Fig.~\ref{fig:autonomous}(e). The non-detected ones continue heating reservoir R up. Here, the detector plays the role of the room reservoir in an absorption refrigerator~\cite{cleuren_cooling_2012,hussein_heat_2016,sanchez_single_2017,erdman_absorption_2018,lu_brownian_2020,manikandan_autonomous_2020,tabatabaei_nonlocal_2022,balduque_quantum_2025} via the heat dumped into it.

Large couplings additionally permit cooling of  reservoir R in the virtual tunneling regime, $\Delta\gg\Omega$. This is shown in Fig.~\ref{fig:autonomous}(b) for $\gamma=10 \Gamma$. The mechanism exploits the localization induced by position measurement and could be named {\it checkpoint cooling}: most particles from L are detected when trying to virtually go through quantum dot C and never reach R. They are rather localized over the barrier and captured by reservoir C, as sketched in Fig.~\ref{fig:autonomous}(f). This way, the few electrons injected from terminal R with energies over $\mu_R$ tend to cool it down, see dark blue regions in Figs.~\ref{fig:autonomous}(b) and \ref{fig:autonomous}(d). They are later also detected and absorbed by C. 

The parameter regime where cooling is observed is similar to what has been predicted for the case of continuously monitored coupled quantum dots~\cite{bhandari2020continuous,elouard2025revealing}.

\section{Measurement Powered Purification}
\label{sec:purif}

\begin{figure}[t]
\includegraphics[width=\linewidth]{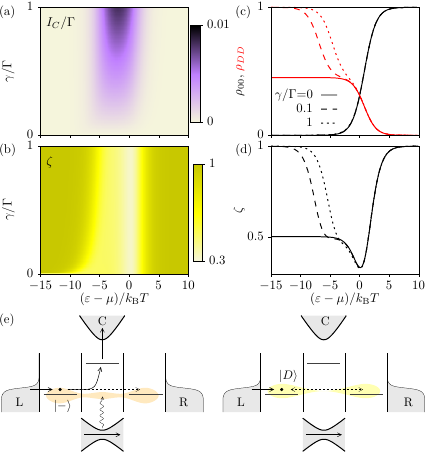}
\caption{\label{fig:3t_purity}\small (a) Zero bias generated current $I_{\rm C}$ and (b) steady state purification as functions of the position of the outer quantum dot levels, $\varepsilon$ and the coupling to the detector. (c) Occupation of the $|0\rangle$ and $|D\rangle$ states and (d) purity for different detector couplings. Parameters are the same as in Fig.~\ref{fig:current_3t_gating}. (e) Scheme of the purification process. }
\end{figure}

The dark state plays an interesting role because it is neither coupled to the central reservoir nor to the detector. As a consequence, if the rates for depopulating it into the left and right reservoirs vanish [i.e., if $W_{\rm 0D}^{\rm L/R}\to0$ for having $f(E_{\rm D}-\mu)\to1$] this state can only be populated. This will be the case when $\varepsilon-\mu\ll-k_{\rm B}T$, which can be easily achieved by tuning the gate voltages~\footnote{It can also be done by reducing the temperature, being careful not to compromise the limit $k_{\rm B}T\gg\Gamma$ where our master equation is valid.}. 
Note that the same applies to the $|-\rangle$ state, which has a smaller energy, as discussed in Sec.~\ref{sec:limits}. However, the detector mixes the $|-\rangle$ and $|+\rangle$ states (by projecting into $|C\rangle=\beta_{-}|-\rangle+\beta_+|+\rangle$), hence introducing a mechanism of depopulating $|-\rangle$, as long as $\varepsilon+\Delta\sim\mu$. 
This way, in the steady state, the system becomes a pure state $\rho\to|D\rangle\langle D|$ by washing all other contributions out via electron tunneling into the reservoirs, as shown in Fig.~\ref{fig:3t_purity}. 
We quantify this effect with the purity of the steady state, which is defined as~\cite{breuer:book}
\begin{equation}
\zeta ={\rm tr}\left(\rho^2\right)=\sum_j\rho_{jj}^2+2(X_{+-}^2+Y_{+-}^2),
\end{equation}
and is, for our four-state system, limited by $0.25\leq\zeta\leq1$.

In the absence of the detector ($\gamma=0$), $I_C=0$ and the state of the system is a mixture of $|-\rangle$ and $|D\rangle$, see Figs.~\ref{fig:3t_purity}(a) and \ref{fig:3t_purity}(c), which is confirmed by a 50\% purity, see Figs.~\ref{fig:3t_purity}(b) and \ref{fig:3t_purity}(d). For finite $\gamma$, $\rho_{DD}\to1$ and $\zeta\to1$ for sufficiently negative $\varepsilon-\mu$. Hence, the density matrix of the system is purified by the detector, producing a dark steady state. Note that this effect will be relevant conditioned on the charge noise-induced dephasing of the dark state~\cite{michaelis_allelectronic_2006} having a rate smaller than $\gamma$.  
For positive gatings ($\varepsilon-\mu\gg k_{\rm B}T$), the purity is trivially 1 because the system can only be empty in that case.

\section{Conclusion}
\label{sec:conc}
In this work, we have demonstrated how quantum measurement backaction, traditionally viewed as a disruptive force, can be harnessed as a thermodynamic resource and a purity generator to power diverse operations in a triple quantum dot (TQD) system. By continuously monitoring the central quantum dot, virtual tunneling events are converted into real occupations, enabling two key functionalities: (i) thermodynamic operations, including power generation, refrigeration, and hybrid energy conversion, and (ii) quantum state purification, where noise from the detector stabilizes the system into a dark state. The TQD setup can also be operated as an autonomous refrigerator, via two different mechanisms: cooling of C is mediated by heat exchange with the detector (analogously to absorption refrigerators), while cooling of R is enabled by the detector avoiding virtual transitions along the TQD for large detection rates (checkpoint cooling). This later mechanism exploits the localization properties of position measurements. The system’s ability to purify itself into a dark state under continuous detection highlights a counterintuitive relation between noise and coherence, where environmental fluctuations suppress decoherence in a symmetry-protected subspace.

These results bridge quantum measurement theory with thermodynamics, showcasing how detection backaction can replace conventional thermodynamic biases (e.g., voltage or thermal gradients) to drive useful operations. The dual functionality of the TQD, as both an engine and a purifier, opens avenues for designing quantum devices that leverage measurement as a fundamental resource. Importantly, the role of tunneling through classically forbidden regions, typically mediated via virtual states, is recontextualized here: measurement backaction does not merely probe but actively shapes these quantum pathways. By injecting energy into the system, the act of measurement allows the particle to transiently occupy virtual states, effectively converting them into real, energetically accessible channels for transport. This tunneling-assisted, measurement-driven activation of transport processes reveals how quantum observation can bridge virtual to real transitions and, in the process, act as a resource for the operation of the device as a measurement engine and a purifier. Future work would explore extending these results to other platforms of interest for quantum technologies (e.g., circuit QED-based devices) and the experimental realization of virtual state-driven measurement engines and state purification in solid-state platforms.

\acknowledgments
RS acknowledges funding from the Spanish Ministerio de Ciencia e Innovaci\'on via grant No. PID2022-142911NB-I00, and through the ``Mar\'{i}a de Maeztu'' Programme for Units of Excellence in R{\&}D CEX2023-001316-M. ANS, BB, and ANJ acknowledge support from the John Templeton Foundation Grant ID 63209. 

\appendix
\section{Vanishing of Heat Current}
\label{app:vanishing}
\begin{figure}[t]
\includegraphics[width=\linewidth]{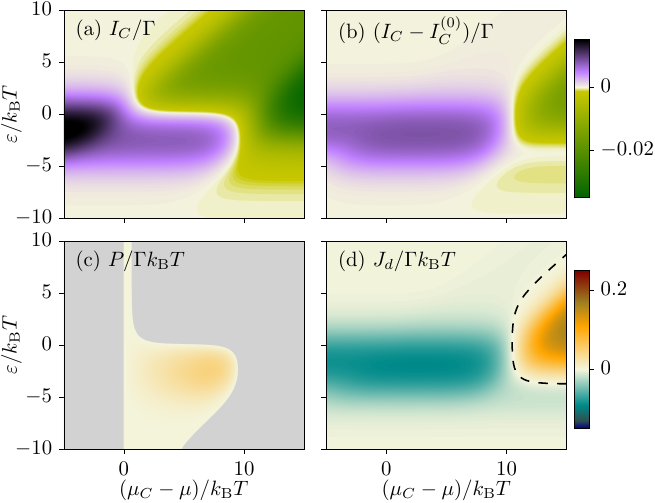}
\caption{\label{fig:3t_vanishinJd}\small (a), Current $I_{\rm C}$ and (b) its modulation with respect to the current in the absence of the detector, $I_{\rm C}^{(0)}$, (c) generated power and (d) heat exchanged with the detector as functions of $\mu_C$ and $\varepsilon$, with $\Omega=\kBT$, $\Delta=10\kBT$, $\Gamma=\gamma=0.1\kBT/\hbar$ and $\mu_L=\mu_R=\mu=0$. Only the region of $P\geq0$ is plotted in (c), with the gray region corresponding to dissipated power. The dashed black line in (d) marks the points where $\rho_{++}=\rho_{--}$.}
\end{figure}

In Fig.~\ref{fig:3t_vanishinJd} we compare the particle transport, the generated power, and the heat exchanged with the detector as functions of the bias $\mu_{\rm C}-\mu$ and the energy of the quantum dots in the regime where $\Delta\gg k_{\rm B}T,\Omega$. A finite current flows into the reservoir C against the bias for $\mu_{\rm C}>\mu$, generating a finite power. Fig.~\ref{fig:3t_vanishinJd}(b) shows the contribution of the detector to the particle current, which vanishes in the same conditions where $J_{\rm d}=0$, cf. Fig.~\ref{fig:3t_vanishinJd}(d).

The vanishing of $J_{\rm d}$ occurs when the system parameters are such that $\rho_{++} = \rho_{--}$, as shown in Fig.~\ref{fig:3t_vanishinJd}(d). When $\varepsilon \lesssim \mu$ (the region of interest where power is generated, see Fig.~\ref{fig:3t_vanishinJd}(c)), this roughly coincides with the chemical potential $\mu_{\rm C}$ crossing the energy $E_+\approx\varepsilon+\Delta$ (see Fig.~\ref{fig:3t_vanishinJd}(d)), i.e., when the occupation of the central dot begins to be occupied by the central reservoir, meaning the occupation becomes real rather than virtual.

As the energy level $\varepsilon$ is further displaced, the influence of the dark state $\ket{D}$ introduces subtleties to the analysis. When $\varepsilon$ lies deep below the Fermi window ($\varepsilon\ll \mu - k_{\rm B}T$), the system becomes trapped in the dark state $\ket{D}$, effectively decoupling from reservoir C. This reflects a dynamical blockade regime where transport is suppressed, see Figs.~\ref{fig:3t_vanishinJd}(a) and \ref{fig:3t_vanishinJd}(d). 

In contrast, when $\varepsilon$ is far above the Fermi window ($\epsilon\gg \mu+k_{\rm B}T$), the system gets populated predominantly via tunneling from reservoir C, as thermal excitations from reservoirs L/R are exponentially suppressed. Here, $|+\rangle\approx|C\rangle$ is populated directly from reservoir C, so the condition $\rho_{++}=\rho_{--}$ depends linerarly on $\varepsilon+\Delta-\mu_{\rm C}$. 
For lower chemical potentials in this region, current flows out of reservoir C through the occupation of state $|-\rangle$ without involving the detector.
In this regime, $\rho_{\rm DD}\to0$, as it is never populated from either L or R.

While these extreme regimes lie outside the operational region of interest ($\epsilon \leq \mu$, where power is generated), they illustrate how the dark state $\ket{D}$ qualitatively alters transport depending on the alignment of $\epsilon$ relative to $\mu$.

\bibliography{biblio}

\end{document}